\documentclass[11pt]{book}
\usepackage{Wiley-AuthoringTemplate}
\usepackage{chapterbib}
\usepackage[sectionbib,authoryear]{natbib}
\usepackage{bm}
\usepackage{nicefrac}
\everymath{\displaystyle}

\begin{document}
\tolerance=10000

\chapter{Kinetic Helicity in the Earth's atmosphere}

\author*[1]{Otto Chkhetiani}
\author[1]{Michael Kurgansky}

\address[1]{
\orgname{A.M. Obukhov Institute of Atmospheric Physics, Russian Academy of Sciences},
\postcode{119017},
     \city{Moscow}, \street{Pyzhyovsky 3}, \country{Russia}}%

\address*{Corresponding Author: Otto Chkhetiani; \email{ochkheti@ifaran.ru}}

\maketitle

\begin{abstract}{}
An overview is given of the helicity of the velocity field (``kinetic'' helicity to distinguish it from the ``magnetic'' helicity used in magnetohydrodynamics, astrophysics, and solar physics; or simply \emph{helicity} in this Chapter) and of the role, which this concept plays in the modern research in atmospheric physics and atmospheric turbulence. General information on helicity is presented including some historical comments and a brief description of the fundamental role that helicity plays in fluid dynamics. Different applications of the helicity concept are discussed to the analysis of various dynamic atmospheric processes, including applications to intense atmospheric vortices, such as tropical cyclones, tornadoes and dust devils, and also to Ekman boundary layer dynamics. Emphasized are helicity balance conditions and the important role that helicity fluxes play in their maintenance. Fundamentals of helical turbulence theory are briefly discussed, and then emphasis is placed on the helical properties of atmospheric turbulence within the atmospheric boundary layer. In particular, the pioneering effort of the research team from the Obukhov Institute of Atmospheric Physics (Moscow) to measure turbulent helicity in natural atmospheric conditions is described.

\end{abstract}


\section{Introduction}\label{sec1}

This Chapter addresses the notion of ``kinetic helicity'', or simply helicity, and its various applications to atmospheric dynamics and atmospheric turbulence. Helicity, or more precisely the helicity bulk density, is defined by a dot product $h=\mathbf{v}\cdot {\bm \omega }$ of the three-dimensional velocity $\mathbf{v}$  and vorticity  ${\bm \omega =}\bm{\nabla}\times\mathbf{v}$  vectors. It is a pseudoscalar quantity, that is, its sign depends on the chosen frame of reference, either right-handed or left-handed, and this sign changes to the opposite under mirror reflection. Helicity is positive when air parcels move along a helical line following the right hand rule. On the contrary, the air motion according to the left hand rule corresponds to negative helicity.

There is a certain hierarchy of atmospheric motions with respect to vorticity: (i) irrotational (potential) flows, (i) rotational flows (vortices), both two-dimensional and three-dimensional but with zero helicity in the latter case, and (iii) helical vortex flows, which are inherently three-dimensional, generally have a complex topology of the vorticity field, and which, in a certain sense, are at the top rung of this hierarchical ladder. In meteorology and atmospheric physics, the interest of many researchers in the concept of helicity is mainly explained by its importance for the study of atmospheric vortex flows, which are usually, but not always, of the convective origin and have a complex structure, sometimes topologically non-trivial. These include tornadoes and dust devils, \citep[e.g.,][]{Tippett2014,Smith2014,Clark2012,Clark2013}; tropical mesoscale vortices \citep{Shenming2011} and tropical cyclones \citep[e.g.,][]{Levina2009,Molinari2010,Yu2011,Levina2019,Chen2021}; and turbulence \citep[e.g.,][]{Chkhetiani2012,Yu2013,Rorai2013,Chkhetiani2018,Vazaeva2017,Vazaeva2021}. Less attention has been paid to the helicity of large-scale extratropical atmospheric motions, explained by atmospheric barocilinicity and Earth's rotation \citep{Etling1985,Kurgansky1990,Bluestein1992}.

This Chapter has the following structure. Section \ref{sec2} provides an overview of helicity, including some historical comments, and presents a brief overview of the fundamental role that helicity plays in fluid dynamics. The different applications of the concept of helicity to the analysis of various dynamic atmospheric processes are discussed in Section \ref{sec3}. The fundamentals of the theory of helical turbulence are briefly discussed in Section \ref{sec4}, and then the helical properties of atmospheric turbulence within the atmospheric boundary layer. In particular, the pioneering effort of the research team of the Obukhov Institute of Atmospheric Physics (Moscow) to measure turbulent helicity under natural atmospheric conditions is described. Final Section \ref{sec5} presents the concluding remarks and formulates the problems that are resisted in the theory of helical turbulence.

\section{Generalities (general information on helicity)}\label{sec2}

In the basic fluid dynamics, the conservation law of helicity for barotropic flows of non-viscous fluids in the presence of only potential external forces was discovered by \citet{Moreau1961} and \citet{Moffatt1969} two centuries after L. Euler presented in 1757 his famous equations of the fluid motion. \citet{Moffatt1969} clarified the topological meaning of helicity as a measure of linkage (knottedness) of vortex lines. This helicity conservation law and its interpretation are closely related to demands of the famous circulation theorem by Lord Kelvin (1868), who as noted in \citet{Moffatt1969} was fully aware of the possibility of the existence of linked and knotted vortex lines. These ideas go back to the classic works of C.F. Gauss on electromagnetism.

Actually, J-J. Moreau [1961] and K. Moffatt [1969] had more immediate predecessors, among them H. Ertel, who in a joint article with C.-G. Rossby \citep{Ertel1949} derived the local conservation law for ideal barotropic flows of a compressible fluid,
\begin{equation} \label{Eqn_1}
\frac{{\rm D}}{{\rm D}t} \left\{\frac{1}{\rho } \bm{\omega }\cdot \left(\mathbf{v}-\bm{\nabla} W\right)\right\}=0,  W=\int\limits_{0}^{t}{\cal L}dt .
\end{equation}
Here, ${{\rm D}/{\rm D}t} $ denotes the total (material) derivative in time. Under the integral sign along the fluid particle trajectory, such that ${\cal L}={{\rm D}W/{\rm D}t} $, stands the fluid dynamical Lagrange function, with $\rho $ as the fluid density, $p$ the pressure and $\Phi $ the gravity potential,
\[{\cal L}=\frac{\mathbf{v}^{2} }{2} -\left(\Phi +\int \frac{dp}{\rho }  \right).\]
The law of conservation of total (integral) helicity follows immediately from \eqref{Eqn_1} if to multiply it by $\rho $ and integrate over the domain occupied by the fluid motion, given that the boundary of the domain is not penetrated by the vortex lines.

Helicity is not a Galilean invariant quantity; see, for example, \citet{Bluestein1992}. There is an analogy with neutrino physics (note how \citet{Moffatt1969} proposed the term ``helicity'' by analogy with the neutrino), where the helicity sign of the neutrino depends on the reference frame in which a neutrino with a non-zero rest mass is observed, that is, moving slower or faster than the neutrino \citep{Goldhaber2011}. The same situation is possible in atmospheric physics. In tornado research, helicity is usually considered in the reference frame associated with the paternal storm, within which a tornado originates. Moving on to another Galilean frame of reference, a situation can be achieved in which, for example, the right rotation of the wind with height (so called in meteorology, because an observer who stands facing the wind discovers that the wind at a higher level blows from the right side, e.g., \citet{Koschmieder1933}), corresponding to positive helicity, will be changed (locally) to left rotation, corresponding to negative helicity.

The concept of helicity proved to be the most powerful and demanded to explain the generation of the magnetic field in an electrically conductive fluid due to the dynamo action \citep{Moffatt1978,Krause1980}. Although nonzero helicity is not necessary in dynamo theory \citep{Gilbert1988}, its presence greatly simplifies the analysis. Based on observational data, \citet{Kuzanyan2000} calculated the helicity distribution in the solar atmosphere, where the helicity determines the alpha effect responsible for generating the magnetic field, for example, \citet{Komm2014}. The rotation of the Sun plays an important role, and it is no coincidence that noticeable (measurable) magnetic fields are found on rotating astronomical objects \citep{Moffatt1978}. \citet{Moiseev1983,Moiseev1988} made a bold attempt to transfer the ideas and methods of dynamo theory to atmospheric physics, but it is not trivial that the last problem presents more conceptual difficulties compared to the first, in particular, in that part which relates to the angular momentum conservation.

In flows with high values of $h=\mathbf{v}\cdot {\bm \omega }$, the vectors $\mathbf{v}$  and ${\bf \omega }$  are almost collinear, so, the Lamb vector ${\bm \omega }\times \mathbf{v}$  in the Euler equations taken in the Gromeka–Lamb form \citep{Landau1987} is almost zero and the cascade of energy to smaller scales is blocked \citep{Speziale1989}. An alternative explanation \citep{Chakraborty2007} is that the complex (non-trivial) topology of the vorticity field effectively hampers the stretching of the vortex lines (tubes) in the bounded space, while stretching is one of main mechanisms of energy transfer along spectrum. The idea is that a knotted vortex line is forced to effectively occupy a larger volume during stretching, which is difficult.

In the incompressible fluid with uniform density $\rho=1$, the integrals of helicity ${\cal H}=\int \mathbf{v}\cdot {\bm \omega }{\rm d}\tau  $, of kinetic energy ${\cal E}={\frac{1}{2}} \int \mathbf{v}^{2} {\rm d}\tau  $, and of enstrophy ${\cal F}={\frac{1}{2}} \int {\bm\omega }^{2} {\rm d}\tau  $, where ${\rm d}\tau $ is the volume element, are related by the Schwartz inequality ${\cal H}^{2} \le 4{\cal EF}$ \citep{Moffatt1969}. When  ${\cal E},{\cal H}={\rm const}$, then  ${\cal F}\ge {\frac{1}{4}} {{\cal H}^{2}/{\cal E}} $ and the inverse energy cascade to large scales is impossible. However, the joint cascade of energy and helicity to small scales is not restricted \citep{Kraichnan1973}. The inequality   becomes the true equality for pure helical motions, known as Beltrami flows, in which ${\bm\omega}\times\mathbf{v}=0$. In atmospheric physics, the concept of Beltrami flow is used with some success to describe the structure of vortices \citep{Kanak1996}. However, it is not still clear whether this concept is just a useful idealization, or whether it corresponds to a motion that really occurs. For example, it has been shown \citep{Kurgansky2013a} that a steady adiabatic Beltrami flow of a fluid with three independent thermodynamic functions of state (moist atmospheric air and/or salt seawater) is dynamically impossible, because its existence would contradict the Ertel vorticity theorem \citep{Ertel1942}.
In Nambu mechanics, which represents a generalization of the Hamiltonian formalism in fluid dynamics \citep{Nambu1973,Nevir1993}, the helicity defined as ${\cal H}={\frac{1}{2}} \int \mathbf{v}\cdot {\bm\omega}{\rm d}\tau  $  has the same level of importance as the energy ${\cal E}$.  Note that with this helicity normalization, which is the most natural for the Nambu theory, the previously written Schwartz inequality acquires a simpler canonical form ${\cal H}^{2} \le {\cal EF}$. Later on, the Nambu formalism was extended to more complex systems of fluid dynamical equation \citep[e.g.,][]{Gassmann2008,Nevir2009,Salazar2010}. The main idea is that the original symmetry properties must be preserved for any approximation of the equations of motion \citep{Salmon2005,Salmon2007}.

Using the Legendre transform, the equation of helicity balance in the Boussinesq fluid can be written in two different forms. The first form uses the potential vorticity \citep{Hide1989}  and the second uses the vertical component of vorticity \citep{Lilly1986,Wu1992}. As an analog of temperature in the general case of a compressible fluid, the geopotential taken with a minus sign appears in the Boussinesq approximation, which considerably facilitates mathematical analysis. The Legendre transform of thermodynamic variables in a compressible fluid provides no such benefit; see also \citet{Kurgansky2006,Kurgansky2017}. Existing attempts to incorporate the concept of helicity into the vortex theory are limited by the Boussinesq approximation, or even by the case of an incompressible fluid with uniform density. An extension to compressible fluid flows awaits its solution.

The definition of helicity does not include the unit of mass. Therefore, the helicity balance equation can be derived without invoking the mass continuity equation. In barotropic fluid flows, helicity is a conserved quantity, when viscosity and non-potential external forces are absent \citep{Moreau1961,Moffatt1969}. In ideal baroclinic fluid flows, non-uniform nonzero values of the potential vorticity determine the source of helicity \citep{Hide1989,Kurgansky1989}. More in detail, a relationship between the helicity bulk density $\mathbf{v}\cdot {\bm\omega}\equiv \mathbf{v}\cdot \bm\nabla \times \mathbf{v}$ and a pseudo-scalar quantity $P={\bm\omega}\cdot \bm\nabla \ln \rho $ was established in \citet{Hide1989} for a generally rotating with the angular velocity $\bm\Omega$  compressible fluid. In the absence of general rotation, $P$  coincides with the potential vorticity for a density-stratified incompressible fluid. That is, \citet{Hide1989} established the relationship
\begin{equation} \label{Eqn_2}
\frac{\partial }{\partial t} \left(\mathbf{v}\cdot {\bm\omega}\right)=-2\frac{p}{\rho } P+\Psi -\bm\nabla \cdot\mathbf{J},
\end{equation}
which is reproduced here in symbolic notation. In equation \eqref{Eqn_2},
\begin{equation} \label{Eqn_3}
\Psi =2\left[{\bm\omega}\cdot \mathbf{F}-\left(2{\bm \Omega }\cdot \mathbf{v}\right)\left(\bm\nabla \cdot \mathbf{v}\right)\right];
\end{equation}
$\mathbf{F}$ denotes the non-potential forces including the frictional force and ponderomotive Ampere force (the latter, for electrically conductive fluids); vector $\mathbf{J}$ is diagnostically expressed in terms of fluid dynamic variables. For a viscous fluid, in the absence of non-potential external body forces, $\mathbf{F}=-\nu \left(\bm\nabla \times {\bm\omega}\right)$, where $\nu $ is the kinematic viscosity. The quantity ${\bm\omega}\cdot \left(\bm\nabla \times {\bm\omega}\right)$ was called ``superhelicity'' in \citet{Hide1989}, and this name seems to have stuck. Although, as R. Hide himself used to say later and as his colleague at the University pointed out to him, it would be purely linguistically more correct to speak of ``hyperhelicity'', since the word ``helicity'' is of Greek origin, and the prefix ``super'', in contrast to the Greek ``hyper'', is of Latin origin. It is worth mentioning that $-2 \nu {\bm\omega}\cdot \left(\bm\nabla \times {\bm\omega}\right)$ describes the time rate of dissipation (destruction) of helicity due to viscosity \citep{Moffatt1969} and superhelicity is not a conserved quantity for inviscid barotropic fluid flows as is helicity itself. Expression \eqref{Eqn_3} includes an additional source of helicity associated with the compressibility of the fluid and its general rotation. However, it was possible to rewrite the equation for helicity balance in a general invariant form \citep{Kurgansky1989}, which includes the Ertel potential vorticity $\Pi _{s} ={\left({\bm\omega}_{\alpha } \cdot \bm\nabla s\right)/\rho } $ for a stratified compressible fluid in a rotating reference frame,
\[
\frac{\partial }{\partial t} \left\{\left({\bm\omega}+4{\bm\Omega }\right)\cdot \mathbf{v}\right\}=-\nabla \cdot \left\{\left({\bm\omega}+4{\bm\Omega }\right){\cal B}+\left({\rm \omega }_{a} \times \mathbf{v}\right)\times \mathbf{v}-T\bm\nabla s\times \mathbf{v}-\mathbf{F}\times \mathbf{v}\right\}
\]
\begin{equation} \label{Eqn_4}
+2T\rho \Pi _{s} +2{\rm \omega }_{a} \cdot \mathbf{F}.
\end{equation}
Here, $s$ is the specific entropy, ${\cal B}=\left({\mathbf{v}^{2}/2} \right)+w+\Phi $ the Bernoulli function, with $w$ as the specific enthalpy and $\Phi $ as the gravity potential, $T$ the absolute temperature in Kelvin, ${\bm\omega}_{a} =\nabla \times \mathbf{v}+2{\bm\Omega }$ the absolute vorticity. At the same time, it was necessary to generalize the concept of helicity to take into account the overall rotation of the fluid,
\begin{equation} \label{Eqn_5}
\mathbf{v}\cdot \bm\nabla \times \mathbf{v}\Rightarrow \mathbf{v}\cdot \left(\bm\nabla \times \mathbf{v}+4{\bm\Omega }\right)\equiv \mathbf{v}\cdot \left({\bm\omega}_{a} +2{\bm\Omega }\right).
\end{equation}
The appearance of a multiplier ``four'' in \eqref{Eqn_5}  is explained by the quadratic nonlinearity of the helicity, with respect to the velocity field. There are different opinions on what quantity should be considered as the absolute helicity. Sometimes, \citep[e.g.][]{Nevir2009}, the scalar product of the absolute velocity and absolute vorticity, $\mathbf{v}_{a} \cdot {\bm\omega}_{a} \equiv \left(\mathbf{v}+{\bm\Omega }\times \mathbf{x}\right)\cdot \left({\bm\omega}+2{\bm\Omega }\right)$, is taken as the absolute helicity, but the presence of the position vector $\mathbf{x}$ in equations written in Euler variables is not always convenient. Sometimes, the absolute helicity is taken as a scalar product of the relative velocity and absolute vorticity,  $\mathbf{v}\cdot {\bm\omega}_{a} =\mathbf{v}\cdot \left({\bm\omega}+2{\bm\Omega }\right)$ \citep{Pichler1998}. However, the latter definition does not fully meet the symmetry requirements and, in particular, with this definition it is not possible to construct the Nambu fluid dynamical brackets \citep[cf.][]{Nevir2009,Salazar2010}. From this perspective, the definition \eqref{Eqn_5} has usage advantages.

\section{Helicity in dynamic atmospheric processes}\label{sec3}

In the baroclinic atmosphere, when the potential vorticity is identically zero,   $\Pi _{s} \equiv 0$, it follows from \eqref{Eqn_4} that the helicity is conserved in an ideal adiabatic case \citep{Kurgansky2006}. Equality to zero, $\Pi _{s} \equiv 0$, is characteristic only of meso- and small-scale atmospheric motions. These flows were extensively studied in the literature, starting from pioneering papers by A. Eliassen (\citet{Eliassen1957,Eliassen1987}; and references therein). Also, based on his helicity balance equation \eqref{Eqn_2} analysis, \citet{Hide1989} argued that vanishing of potential vorticity is of general validity for both natural and laboratory convection. For large-scale extratropical atmospheric processes, the vector ${\bm\omega}_{\alpha } $ ``permeates'' the isentropic surfaces at a non-zero angle. Therefore, non-zero values  $\Pi _{s}$  and thus non-zero helicity sources always exist. For large-scale, ``2$\nicefrac{1}{2}$-dimensional'' processes \citep[cf.][]{Kurgansky2002}, the helicity quantifies the “non-self-similarity” of the velocity field at different height levels, when, for example, the change of the wind direction with height occurs due to the ”thermal wind” effect. Helicity vanishes in barotropic or equivalent-barotropic atmospheric motions. The significant change of wind direction with height, and the accompanying helicity concentration, occurs in the Ekman boundary layer. This is the famous Ekman spiral with the right rotation of the wind with height (a veering wind that turns clockwise with height) and positive helicity in the main bulk of the Ekman boundary layer in the Northern Hemisphere and left wind rotation (a backing wind that turns counterclockwise with height) and negative helicity in the Southern Hemisphere.

\citet{Etling1985} and \citet{Lilly1986} introduced the concept of helicity into atmospheric physics. In the decade that followed, only a limited number of articles devoted to “atmospheric helicity” were  published. Then the number of articles of this type began to increase steadily, especially those dedicated to rotating supercells and tornadoes and applications of the concept of the storm relative environmental helicity \citep{Droegemeier1993} and/or storm-relative helicity \citep{Markowski1998} \. These helicities, which are measured in the frame of reference that moves with the parental storm, are one of main predictors of tornadogenesis, see, for example, \citet{Doswell2006}. To avoid misunderstanding, we note that in fluid dynamics the term ``relative helicity'' usually refers to  $\chi ={\left(\mathbf{v}\cdot {\bm\omega}\right)/\left(\left|\mathbf{v}\right|\cdot \left|{\bm\omega}\right|\right)} $, that is, to the cosine of the angle between the vectors  $\mathbf{v}$ and ${\bm\omega}=(\bm\nabla)\times\mathbf{v}$. It can be stated that helicity is certainly useful and in great demand in the meso and micrometeorology. We note in this regard that \citet{Han2006} introduced the ``shearing wind helicity'' and ``thermal wind helicity'', which differ from the commonly used helicity. According to \citet{Han2006}, these newly introduced helicities are useful in the analysis of circulation systems and have been applied to the diagnosis of Hurricane Andrew (1992). As a whole, we can agree that helicity is a useful diagnostic for identifying certain types of long-lived storms, and the more data is available and evoked to characterize these storms, the better.

In general, helicity is of indisputable importance for the diagnosis of various atmospheric processes, but does it have predictive value? \citet{Levina2010,Levina2011}, see also \citet{Levina2019}, showed that (integral) helicity is an important indicator of development of the large-scale instability leading to the tropical cyclogenesis. Using the output data from the Eta regional atmospheric model, \citet{Lavrova2010} found that the calculated helicity can serve as one of the earliest predictors of the genesis of Mediterranean cyclones. The development of this research direction is of undoubted interest, in particular in the part that refers to the relationship between anomalous “atmospheric helicity” and the further development of weather systems \citep[cf.][]{Vazaeva2021}.

\citet{Hide1989,Kurgansky1989,Kurgansky1993,Kurgansky2002,Zhemin1994,Chkhetiani2001,Ponomarev2003,Ponomarev2005,Deusebio2014,Chkhetiani2018}, and most recently in \citet{Vazaeva2021}, investigated  different dynamic aspects of helicity in the Ekman boundary layer. In particular, \citet{Zhemin1994} took into account the non-linearity of the equations of motion and revealed the cyclone/anticyclone asymmetry in the helicity values in the Ekman boundary layer. The influence of baroclinicity (thermal wind) was also considered by \citet{Zhemin1994}, but contrary to \citet{Kurgansky1989,Kurgansky1993,Kurgansky2002}, the helicity balance in the Ekman boundary layer was not analyzed by these authors, which, in virtue of additional effects indicated in \citet{Zhemin1994}, is an interesting problem for future studies \citep[cf. also][]{Deusebio2014,Kurgansky2017,Chkhetiani2018}.

Helicity density has a dimension of acceleration, and its proper reference value on Earth is the acceleration due to gravity  $g=9.81\, {\rm m}\, {\rm s}^{-2}$. Helicity quantifies the three-dimensional organization (e.g. \citet{Levich1985}, as well as Table \ref{tabl-1} adapted from \citet{Cieszelski1999} who gave the summary of estimates in \citet{Etling1985} and \citet{Zhemin1994}; these were further extended and refined in \citet{Vazaeva2021}) and provides stability and durability to the atmospheric vortex structures that possess helicity \citep{Lilly1986}.

\begin{table}
\caption{Scale and helicity bulk density of different atmospheric vortex structures (\citet{Cieszelski1999}; cf. \citet{Kurgansky2017} and \citet{Vazaeva2021})\label{tabl-1}}
{%
\begin{tabular}{@{}cccc@{}}
\toprule
Vortex structure    &Height, m    &Radius, m   &Helicity, m/s${}^{2}$\\
\midrule
Tropical hurricane & 10${}^{4}$ & 10${}^{5}$ & 10${}^{-}$${}^{1}$ \\
Storm & 10${}^{4}$ & 10${}^{3}$ & 10${}^{-}$${}^{2}$ \\
Tornado & 10${}^{3}$ & 10${}^{2}$ & 10${}^{1}$ \\
Ekman layer & 10${}^{3}$ & $-$ & 10${}^{-}$${}^{1}$ \\
Rotating thermal & 10${}^{3}$ & 10${}^{2}$ & 10${}^{-}$${}^{2}$ \\
Dust devil & 10${}^{2}$$-$10${}^{3}$ & 10${}^{1}$ & 10${}^{1}$ \\
\botrule
\end{tabular}}
{}
\end{table}

In addition to the vortices indicated in Table \ref{tabl-1}, high values of $\left|h\right|$ are characteristic of, e.g., horizontal vortex structures in the planetary boundary layer: rolls and ``cloud streets'' \citep{Etling1985}. For vertical columnar vortices in approximate cyclostrophic equilibrium, the most convenient estimate of the helicity bulk density is $\left|h\right|\sim {VW/R} $, where $V$ is the characteristic azimuthal velocity, $W$ the characteristic vertical velocity and $R$ the vortex core radius. More precisely, the contribution of the azimuthal velocity and vorticity components to helicity is estimated as $\left({2V/R}\right)W\pi R^{2} H$, where $H$ is the vortex height. The contribution to the helicity of the radial components of the velocity and voricity is usually negligible, but the contribution of the vertical components turns out to be approximately equal to the contribution of the azimuthal components calculated above, so that the total helicity of the vortex is estimated as ${\cal H}=4\pi VWRH$, which, in contrast to the kinetic energy of a vortex, can be quite finite as, for example, it occurs for a combined Rankine vortex. This is one more virtue of the helicity concept. It is possible to write equivalently that ${\cal H}=2\cdot 2\pi RV\cdot WH$ \citep[cf.][]{Kurgansky2013b}, which is quite consistent with the Moffat's [1978] result for the value of total helicity in the case of entangled toroidal and poloidal circulations. For mesoscale tropical cyclones, the contribution to the helicity of toroidal (azimuthal) components of $\mathbf{v}$ and $\bm\omega$ is much greater than the contribution of poloidal components, associated with the so-called secondary, or meridional, circulation; the former contribution has the order of magnitude of $10^{-1} {\rm m}\, {\rm s}^{-2} $, while the latter is $\sim 10^{-3} {\rm m}\, {\rm s}^{-2} $ \citep[cf.][]{Cieszelski1999,Kurgansky2017}. For large-scale quasi-geostrophic atmospheric processes, as the scale estimation shows by taking into account that these motions are quasi-geostrophic, the contribution of the horizontal components of  $\mathbf{v}$ and $\bm\omega$ to the helicity is approximately five orders of magnitude greater than that of the vertical components \citep[cf.][]{Kurgansky2017}.

Representation of helicity $h=\mathbf{v}\cdot{\bm\omega}$  as a sum of three contributions corresponding to three (locally) orthogonal directions in space, when each is equal to the product of the corresponding components of velocity and vorticity, is not unique. Equally useful is the alternate representation \citep{Hide1975}, also as a sum of the three terms but arranged differently. For each term a direction in space is indicated and only those components of velocity that are in the plane orthogonal to this direction are considered. In this way, in practically important cases, the main contributors to helicity are grouped into a single term but are not dispersed in different contributions, as occurs with the first, more traditional representation. For example, in the Ekman boundary layer, they are grouped into a term that corresponds to the vertical direction and the horizontal velocity components (and their vertical derivatives). For vertical convective vortices, such as tornadoes and dust devils, they are grouped into a term that corresponds to the radial direction and the azimuthal and vertical velocity (and their radial derivatives).

\begin{figure}
 \includegraphics[width=1.0\textwidth]{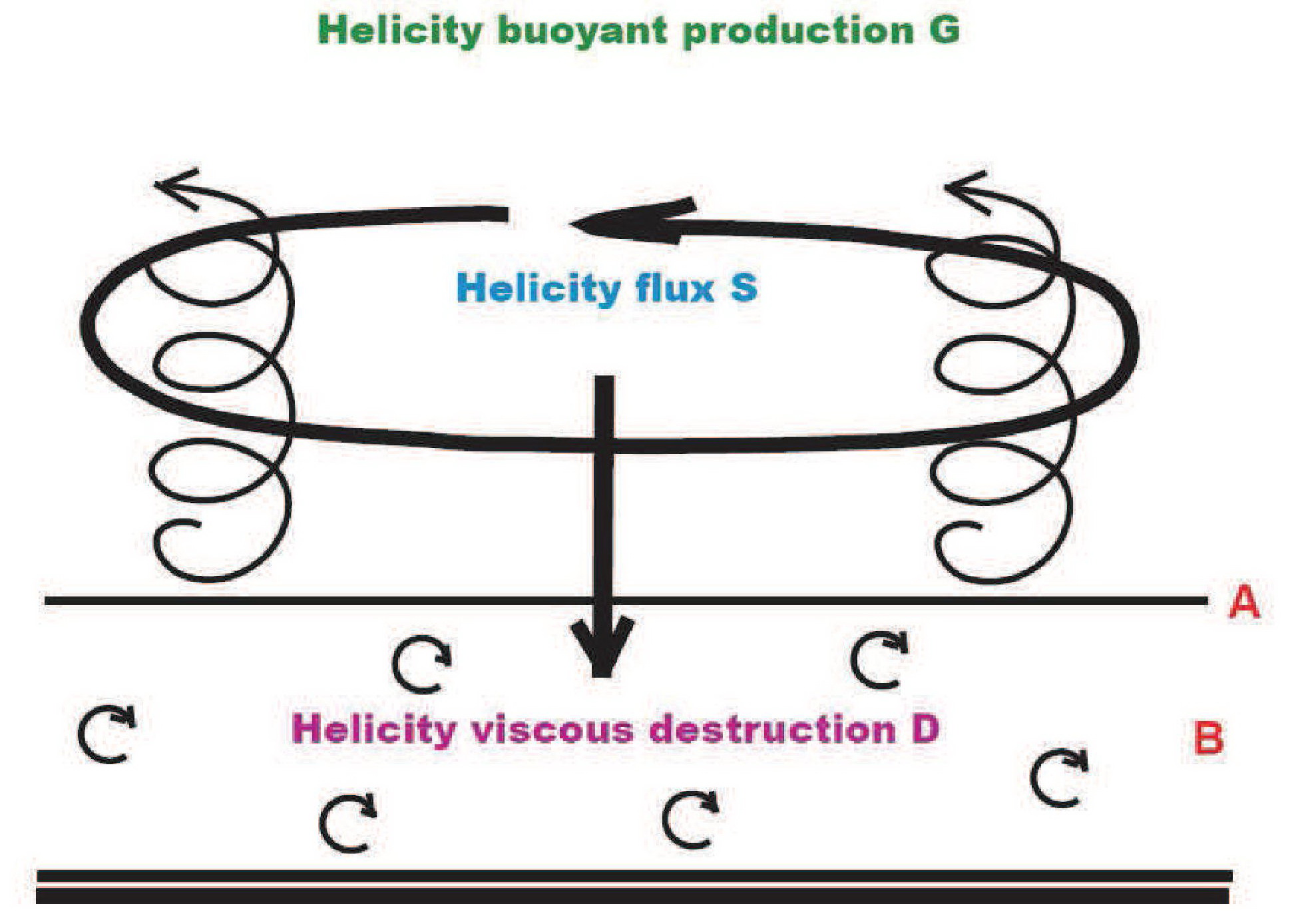}
\caption{A schematic of helicity balance in intense atmospheric vortices. The vortex flow is outlined by a thick curved arrow. Helicity flux $S$  is shown by a downward pointing arrow. Small-scale turbulence within the boundary layer is presented by small curved arrows. A thick double horizontal line shows the Earth’s surface.
\label{Fig1}}
\end{figure}

In the atmosphere, the helicity generation and destruction regions do not usually coincide, but are generally separated in space. Therefore, the helicity fluxes are important to maintain the helicity balance. A scheme of helicity balance in intense atmospheric vortices is presented in Fig. \ref{Fig1}. It is considered a steady axisymmetric vortex. Most of the helicity production $G$ due to vertical convection in the presence of overall rotation (shown schematically by vertically oriented helical lines), occurs above a horizontal surface $A$, which marks the top of a surface boundary layer $B$. The helicity destruction $D$, due to small-scale turbulent viscosity, is limited to $B$. Under real circumstances, it is optimal to take $A$ at a height of several meters to a few tens of meters above ground level.

 \citet{Kurgansky2008} proposed using vertical helical flux in small and mesoscale vortices as a measure of their intensity. For a steady axisymmetric vortex in the polar cylindrical coordinates under the neglect of the small radial velocity component $u$, this downward directed helicity flux through the upper level $ A$ of the turbulent boundary layer reads \citet{Kurgansky2017}
\begin{equation} \label{Eqn_6}
S=\iint\limits _{A }\left(2v^{2} -w^{2} \right)\,  \omega _{z} {\rm d}\sigma .
\end{equation}
Here, $v$ is the azimuthal velocity component, $w$ the vertical velocity component, $\omega _{z} $ the vertical vorticity, and ${\rm d}\sigma $ the surface area element; it has been used that $rv\left. w^{2} \right|_{r=0}^{r=\infty } =0$.  Here we come to an important point. As already follows from \citet{Moffatt1969} and  confirmed by the analysis of \eqref{Eqn_4}, the helicity flux has both the advective (velocity-driven) and non-advective (along the vortex lines) component. As a rule, the latter component, for example in intense atmospheric vortices, significantly exceeds the former, as it is accounted for in \eqref{Eqn_6}; see also \citet{Kurgansky2017}. The effect of Earth rotation with the angular velocity ${\bm\Omega }$ has been neglected. If needed, its account would result in appearance of the additional term $\iint\limits _{A } \left(2{\bm\Omega }\cdot \mathbf{k}\right)\,  v^{2} {\rm d}\sigma $ on the right side of \eqref{Eqn_6}, where $\mathbf{k}$ is the unit vertical vector.

In atmospheric vortices of small and mesoscales, as a rule, $u<\left(\ll \right)w<\left(\ll \right)v$. For example, based on direct measurements from aircraft in waterspouts in \citet{Leverson1977}, the characteristic values are as follows: $v=25\, {\rm m}\, {\rm s}^{-1} $, $w=8\, {\rm m}\, {\rm s}^{-1} $ and $u=1\, {\rm m}\, {\rm s}^{-1} $. Therefore, instead of \eqref{Eqn_6}, the approximate formula can be used, see \citet{Kurgansky2008,Kurgansky2017}
\begin{equation} \label{Eqn_7}
S\cong \iint\limits _{A }2v^{2} \omega _{z} {\rm d}\sigma  ,
\end{equation}
or alternatively, $S\cong \left({8\pi /3} \right)\int\limits_{0}^{\infty }v^{3}  {\rm d}r$, if only $v^{3} r\to 0$ with $r\to \infty $, where $r$ is the radial coordinate. \citet{Glebova2009} applied the method of helicity fluxes to the Man-Yi (2007) and Wilma (2005) tropical cyclones and \citet{Farahani2017} did it for the tropical storm Gonu (2007).

For the combined Rankine vortex with the uniform vertical flow inside the vortex core and under neglect of the Coriolis force, the vertical helicity flux in a vortex is estimated by the value $S=2\pi V\left(V^{2} -W^{2} \right)R$, see \eqref{Eqn_6}, where $V$ is the maximum azimuthal velocity, $W$ the vertical velocity and $R$ the vortex core radius, and if $V\gg W$, then ${\rm S}\approx 2\pi V^{3} R$; see \eqref{Eqn_7} and \citet{Kurgansky2013b}. The natural time scale for vortices is the time of their complete revolution around the axis ${2\pi R/V} $. For example, the use of data in \citet{Lorenz2013} allows us to conclude that dry-convective small-scale dust-devil vortices, regardless of their size, have a life-duration on average about 60 revolutions. Simply, in large vortices this revolution takes a longer time, therefore, in absolute units of time, they live longer. The ratio ${{\cal H}/S} $, where ${\cal H}=4\pi VWRH$ is the total helicity of the vortex flow and $H$ is the height of the vortex  (see above), determines the characteristic time of the helicity transformation (helicity metabolism) in the vortex flow. For dust devils, this time is short and equates to only few complete revolutions of the vortex around its axis. If we put ${H/R} =10$ \citep{Lorenz2013,Jackson2020} and accept that ${V/W} =3$, then the estimate shows that the time of the complete renewal of helicity in the vortex is $\sim 6$ complete revolutions of the dust devil around its axis, i.e. during a typical dust-devil vortex lifetime, the vortex helicity has time to be completely renewed about 10 times. From this perspective, one can say that helicity is a kind of constantly renewing corset (frame) that energy fills in. The claim  about a shorter helicity destruction time, compared to energy dissipation, is generally consistent with the conclusions of \citet{Levshin2013}, who showed that  helicity in homogeneous turbulence decays faster than energy. The same is true for the large-scale extratropical atmospheric motions. The rate of destruction of the helicity within the Ekman boundary layer is equal to the downward flux of helicity from the free atmosphere. This flux is determined (per unit area of the Earth’s surface) by the product of the Coriolis parameter and the square of the geostrophic wind speed in the free atmosphere \citep{Kurgansky1989,Kurgansky1993,Kurgansky2002}; also \citet{Deusebio2014}. Therefore, the helicity transformation time in large-scale extratropical atmospheric processes is given by the reciprocal of the Coriolis parameter ($\sim 10^{4} \, {\rm s}$) and by a factor of ten it is less than the characteristic duration of large-scale cyclones and anticyclones.

Until now, the main emphasis has been on observations of helicity but not on its dynamics. The above arguments about helicity balance and  its “metabolism” apply to a quasi-steady mature stage of atmospheric vortices. The issue of helicity generation (genesis) is more complex and non-trivial. Helicity is conserved and therefore cannot be generated in barotropic flows of inviscid fluids in the presence of potential forces \citep{Moreau1961,Moffatt1969}. So, the genesis of helicity requires the violation of the above conditions, that is, it requires the joint action of atmospheric baroclinicity, which often takes the form of vertical convection leading to vertical flows, and the general rotation of the flow system, either directly or indirectly related to the Earth’s rotation. In the latter case the sense of Earth’s rotation is “transmitted” through the sequence of emerging circulation systems of progressively decreasing spatial scale (cold atmospheric fronts, mesocyclones, etc.), in which tornadoes are embedded, almost all them (with rare exceptions) being cyclonic vortices. There are indications that surface friction may also contribute to the genesis of helicity. The small-scale dynamics of helicity is important for tornadoes. In particular, these are Hasimoto solitons \citep{Hasimoto1972} propagating along strong vortex filaments (thin tornadoes) and \citet{Aref1984} have provided photographs of a tornado near Braman, Oklahoma (11  May 1978) showing localized large-amplitude helical twist in the vortex core (a “soliton”).  \citet{Holm2004} showed that the nonlinear Schrödinger equation in \citet{Hasimoto1972} is replaced by the integrable complex modified Korteweg–de Vries equation having soliton solutions, when the Hamiltonian in the underlying theory is not the kinetic energy but helicity. The classical case of kinetic energy and the case of helicity discussed in \citet{Holm2004} exhaust all Hamiltonians that have fluid dynamical significance (ibid).

\section{Helical turbulence in the Earth's atmosphere}\label{sec4}

Moving from the discussion of general principles in Section \ref{sec2} to a more detailed consideration of helical turbulence, particularly in the Earth’s atmosphere, we first mention successful attempts to numerically reproduce cascades of energy and helicity directed differently, but taking into account the general rotation of the fluid \citep{Mininni2010a,Mininni2010b,Pouquet2010}. However, \citet{Biferale2012,Biferale2013} and \citet{Sahoo2017} showed that without general rotation but by allowing interactions between helical modes of the same sign (homochiral modes), the inverse cascade of energy is possible as in the two-dimensional turbulence. \citet{Gledzer2015,Chkhetiani2017}, also using the turbulent cascade model with conservation of helicity \citep{Gledzer1973}, show that to reproduce the inverse energy cascade from small-scale disturbances, a certain level of “helical noise” is necessary in large-scale modes, which is generated by external forcing. These conclusions are also supported by an analysis of second-order moments in the quasi-normal approximation. Similar results were also obtained in \citet{Stepanov2015}. Another line of research uses the formulation of three-dimensional turbulent fields in terms of helical modes introduced in  \citet{Knorr1990} and  \citet{Waleffe1992}, which was applied in \citet{Rathmann2017} to investigate some details in joined energy and helicity cascades.

An important research line is related to direct numerical simulation (DNS) studies on helicity which are only briefly described here. In particular, they include DNS studies on turbulence decay by  \citet{Holm2002}. Their diagnostics showed that the turbulence initial value problem evolves through an early stage, with still negligible dissipation, when vortex sheets form from smooth initial conditions and then coil into helical vortex tubes. Once the helical vortex tubes have formed, viscous dissipation grows rapidly in the DNS and the energy spectrum gradually approaches the classical “-5/3” law characteristic of a turbulent energy cascade, but before is achieved, the energy spectrum remains steeper than “-5/3”; see Fig. 6 in  \citet{Holm2002}. Once a clear “-5/3” spectrum appears, the energy decay becomes self-similar. The numerical study of the joint cascade of energy and helicity by three-dimensional DNS was carried out by  \citet{Chen2003} and verified their theoretical conclusions that the Kolmogorov microscale of turbulence is the minimum scale of helical motions for a joint cascade of energy and helicity (see also below). DNS studies of helical turbulent flows continue and are actively developing today. It is beyond the scope of this review to list all relevant papers; see, for example, a recent article by  \citet{Biferale2019}.

\citet{Chkhetiani1996} obtained an important relation between the two-point triple correlation of the velocity and the rate of helicity transfer over the spectrum and, therefore, of its viscous destruction at small scales. Later it was called the ``2/15 law'' for helicity, which is dual to the ``4/5 law'' of A.N. Kolmogorov for the velocity field in three-dimensional non-helical turbulence, and is similar to the ``4/3 law'' of A.M. Yaglom for a passive scalar \citep{Chkhetiani2008}; see also \citep[{\S} 6.1.6]{Golitsyn2012}. The ``2/15 law'' allows to obtain the spectra of helicity, both the ``-5/3 law'', when the helicity behaves as a passive scalar, and the more hypothetical ``-4/3 law'', when the helicity cascade is decisive.

 More in detail, two cases of the turbulent cascade of helicity are possible. The first case is the joint cascade of energy and helicity towards small scales. The spectral slope of helicity coincides in this case with the slope of the energy spectrum and is equal to $-$5/3
\begin{equation} \label{Eqn_8}
\hat{E}(k)=c_{E} \varepsilon ^{2/3} k^{-5/3} ,\quad \hat{H}(k)=c_{H} \eta \, \varepsilon ^{-1/3} k^{-5/3} .
\end{equation}
The second case is the direct cascade of helicity and inverse cascade of energy (Brissaud et al. 1973)
\begin{equation} \label{Eqn_9}
\hat{E}(k)=c_{E}^{*} \eta ^{2/3} k^{-7/3} ,\hat{H}(k)=c_{H}^{*} \eta ^{2/3} k^{-4/3} .
\end{equation}
Here $\hat{E}(k)$ and $\hat{H}(k)$ are the spectral densities of energy ${\cal E}={\frac{1}{2}} \left\langle \mathbf{v}^{2} \right\rangle $ and helicity ${\cal H}=\left\langle \mathbf{v}\cdot {\bm\omega}\right\rangle $ , respectively; $\varepsilon =\nu \langle (\bm\nabla \times \mathbf{v})^{2} \rangle $ is the specific kinetic energy dissipation rate and $\eta =\nu \langle {\bm\omega}\cdot \bm\nabla \times {\bm\omega}\rangle $ is the time rate of helicity viscous destruction; $k$ is the wavenumber, $\nu $ the kinematic viscosity and angular brackets denote ensemble averaging. Additionally, we note that in this case, the Obukhov--Corrsin spectrum of temperature fluctuations (passive scalar) and the co-spectrum of the temperature $T$ and vertical velocity $U_{3} $ fluctuations will also have different slopes \citep{Moiseev1996}
\begin{equation} \label{Eqn_10}
\hat{E}_{TT} (k)\sim N\eta ^{-1/3} k^{-4/3} ,   \hat{E}_{TU_{3} } (k)\sim N\eta ^{-1/6} k^{-7/6}
\end{equation}
compared with the $-$5/3 in the usual case. Here, $N=\chi \left\langle (\nabla T)^{2} \right\rangle $ is the average rate of dissipation of temperature inhomogeneities. Let us also note that the analogue of the Bolgiano--Obukhov scale $L_{*} =\eta ^{-5} N^{3} (g\beta )^{6} $ arises in comparison with $L_{BO} =\varepsilon ^{5/4} (g\beta )^{-3/2} N^{-3/4} $, where $\beta ={1/T_{0} } $ with $T_{0} $ as the constant reference temperature. For a long time, the existence of the helical cascade \eqref{Eqn_9} was considered hypothetical. In particular, the conditions for the realization of inverse energy cascades were considered difficult to fulfill in real conditions, although, certain spectra observed in the atmosphere and in laboratory experiments had spectral slopes quite close to those in \eqref{Eqn_9} \citep{Chkhetiani2006}. It is only relatively recently that such regimes began to be reproduced in numerical experiments using helicity pumping at different scales \citep{Biferale2012,Biferale2013,Sahoo2017,Gledzer2015}.

The presence of helicity in turbulence manifests itself in the appearance in the tensor of two-point triple correlations of the velocity field those components, which are identically equal to zero in its absence (Chkhetiani et al. 1996), and it holds that
\begin{equation} \label{Eqn_11}
\langle \delta \mathbf{v}_{l} (x|r)\cdot [\mathbf{v}_{t} (x+r)\times \mathbf{v}_{t} (x)]\rangle =\frac{2}{15} \eta r^{2} .
\end{equation}
The indicated dependence (the so-called 2/15 law) is similar to the Kolmogorov 4/5 law for longitudinal velocity correlations \citep[cf.][]{Monin19711975}
\begin{equation} \label{Eqn_12}
\langle \left(\delta \mathbf{v}_{l} (x|r)\right)^{3} \rangle =-\frac{4}{5} \varepsilon r.
\end{equation}
and is the exact result following from the Navier--Stokes equations. Here, the velocity field is divided into longitudinal and transverse components respective to the vector $r$ connecting the observation points, $\mathbf{v}_{l} =\left(\mathbf{v}\mathbf{r}\right)\mathbf{r}/r^{2} $, $\mathbf{v}_{t} =\mathbf{v}-\mathbf{v}_{l} $, $\delta \mathbf{v}_{l} \left(\mathbf{x}|\mathbf{r}\right)=\left(v_{l} (\mathbf{x}+\mathbf{r})-v_{l} (\mathbf{x})\right)\mathbf{r}/r$. Denoting $r_{i} =r\delta _{ix} $ we obtain for the two-point triple tensor of velocity and vorticity correlations the following representation \citep{Chkhetiani2008}
\begin{equation} \label{Eqn_13}
T_{i,j,k} =\left\langle v_{i} \left(\mathbf{x}\right)v_{j} \left(\mathbf{x}\right)\omega _{k} \left(\mathbf{x}+\mathbf{r}\right)\right\rangle =\frac{\eta r}{15} \delta _{ij} \delta _{kx} -\frac{\eta r}{10} \left(\delta _{ix} \delta _{jk} +\delta _{ik} \delta _{jx} \right).
\end{equation}
The understanding of the presence of helicity in atmospheric turbulence came later. Thus, in the work of \citet{Novikov1972}, where the statistical properties of the vorticity flux tensor
\begin{equation} \label{Eqn_14}
\langle v_{i} \omega _{j} \rangle =\frac{\varepsilon _{ikl} }{4\pi } \int\limits_{V} \frac{r_{l} }{r^{3} } \langle \omega _{k} (\mathbf{x}')\omega _{j} (\mathbf{x})\rangle {\rm d}^{3} \mathbf{x}',
\end{equation}
were discussed, the diagonal elements of this tensor, which are directly related to helicity, were not considered. According to \citet{Novikov1972}, only four non-diagonal elements of \eqref{Eqn_14}, $\left\langle u'\omega '_{2} \right\rangle $, $\left\langle v'\omega '_{1} \right\rangle $, $\left\langle w'\omega '_{1} \right\rangle $ and $\left\langle u'\omega '_{3} \right\rangle $, where $\mathbf{v}=\left(u,v,w\right)$, should be different from zero in the atmospheric boundary layer (ABL) turbulence. The first measurements of the covariance between the velocity and vorticity components and the corresponding co-spectra and spectra, were carried out at heights of 1.2 and 5 m at the Tsimlyansk test site of the Institute of Atmospheric Physics, Russian Academy of Sciences \citep{Koprov1988} using an acoustic circulation-meter (the so-called, ``circulimeter'') first designed and manufactured by \citet{Bovsheverov1971}. The observed differences from zero of the covariance, associated with the circulation in the vertical plane parallel to the mean flow directed along the $x$-axis and corresponding to $\left\langle v'\omega '_{2} \right\rangle $, were clearly not consistent with the conclusions of \citet{Novikov1972}. In addition, a strong correlation was found between the velocity circulation and the temperature field.

\begin{figure}
 \includegraphics[width=1.0\textwidth]{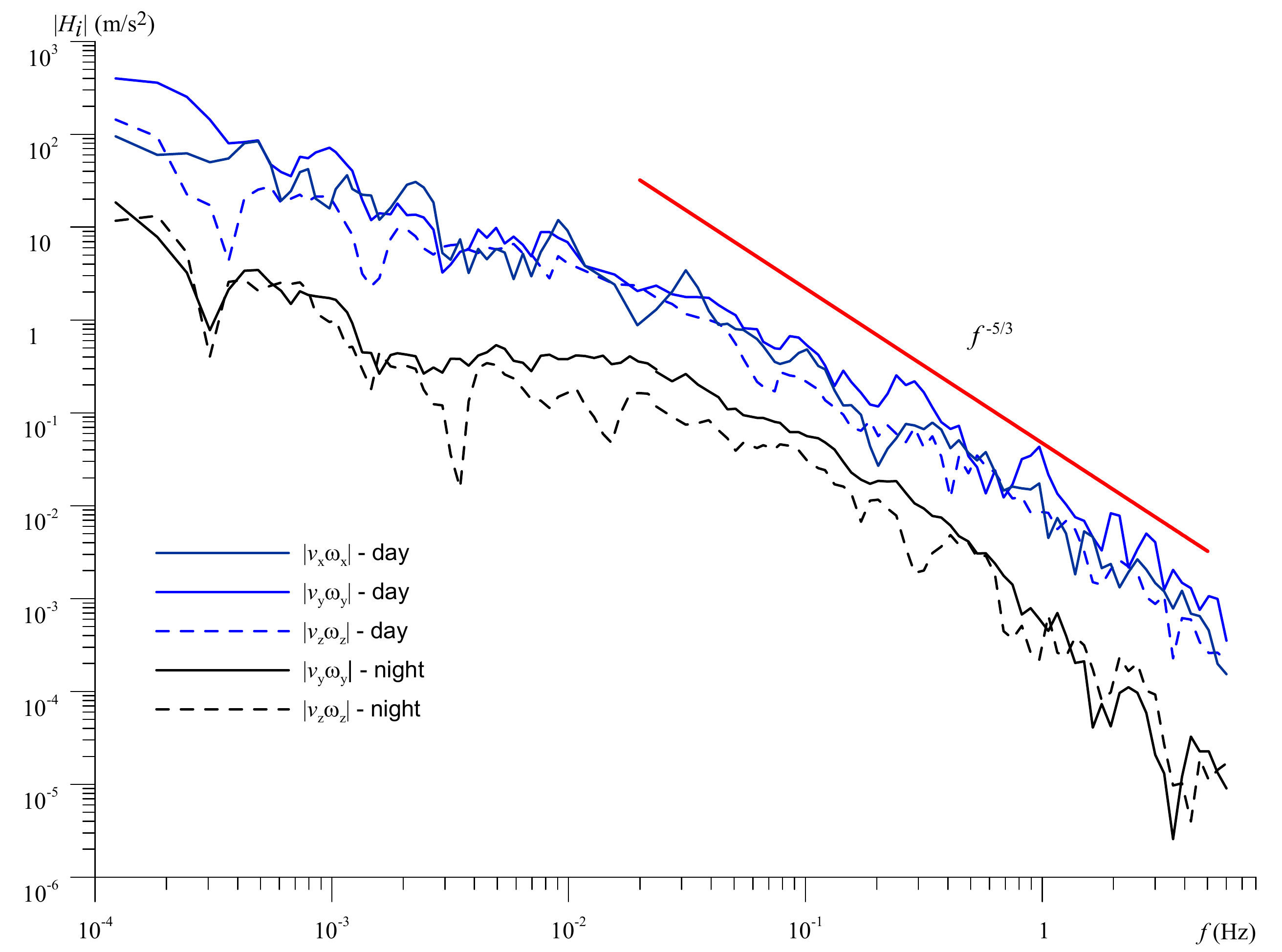}
\caption{Frequency spectra of the contributions to helicity calculated based on the 2-hour records: the nighttime of 11.09.14 with the onset at 22:17 local time and the daytime of 12.09.14 with the onset at 13:41 local time \citep{Koprov2005}.
\label{Fig2}}
\end{figure}

\begin{figure}
 \includegraphics[width=1.0\textwidth]{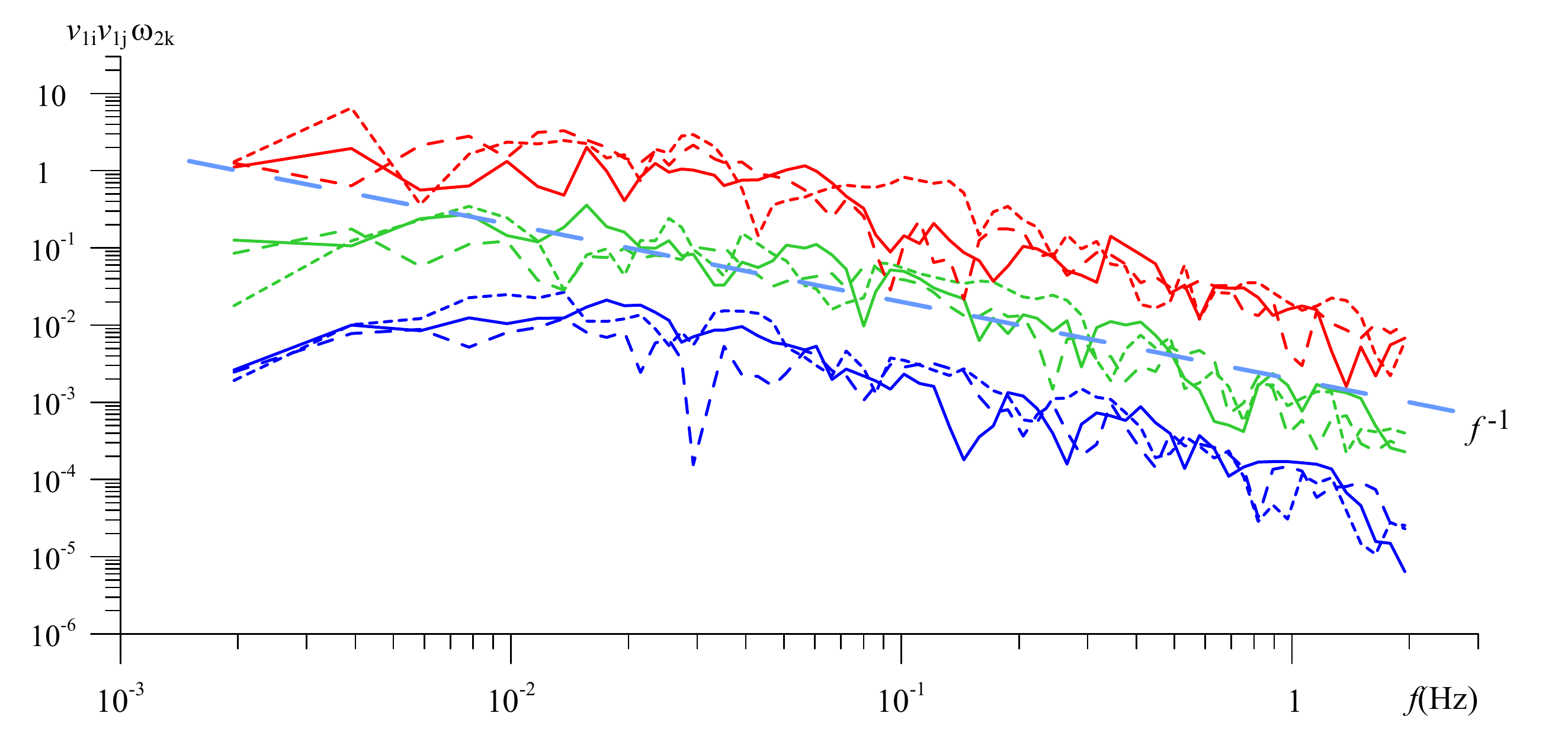}
\caption{The triple correlation frequency spectra calculated from the 2-hour records of the velocity and circulation components  from 28.08.04 with the onset at 11:38 local time (red line), at 16:41 local time (green line), and from 11.09.04 with the onset at 22:17 local time (blue line). The solid line shows the $\left\langle v_{y} v_{z} \omega _{y} \right\rangle $ spectrum, the long dashed line -- $\left\langle v_{x} v_{z} \omega _{y} \right\rangle $ spectrum, and the short dashed line -- $\left\langle v_{x} v_{y} \omega _{y} \right\rangle $ spectrum.
\label{Fig3}}
\end{figure}

To measure  helicity, measurements of all three components of the velocity and vorticity are required. Vorticity measurement techniques can be conditionally divided into direct and indirect ones. Direct measurements with hot-wire arrays were carried out in \citet{Wyngaard1969,Antonia1988,Vukoslavcevic1996} and \citet{Kholmyansky2001}. Providing the ability to carry out measurements on scales close to the dissipation range, this technique poorly matches to the conditions of a real ABL. Indirect methods are associated with the measurement of circulation along a certain contour and, accordingly, the determination of the average vorticity over the area encircled by the contour, based on the application of the Stokes theorem \citep{Bovsheverov1971,Koprov1988,Koprov1994,Koprov2005}. Ideologically close techniques for measuring atmospheric circulation were also proposed in \citet{Jordan1980,Ohtou1983} and \citet{Mitsuta1984}. Such techniques were often used to determine the characteristics of oceanic turbulence \citep{Rossby1975,Gaynor1977,Thwaites1995,Sanford1999}. The procedure for calculating the velocity circulation was discussed in detail by \citet{Longuet-Higgins1982}. In \citet{Koprov1994,Koprov2005}, synchronized measurements of velocity and circulation were made using one- and two-component acoustic ``circulimeters'' with a contour diameter of 1 m together with an acoustic anemometer. All three sensors were placed on top of a 46-m high mast, with the anemometer in the center, and the ``circulimeters'' on both sides of it at a distance of 0.8 m. The spectra of helicity components with a slope close to --5/3 were observed (Fig. \ref{Fig2}); see \citet{Koprov2005}. An integral estimate of helicity from the observed spectrum gives the values of 0.02--0.03 m${}_{ }$s${}^{-}$${}^{2}$. For the components of the two-point tensor triple product of the velocity and vorticity components $\left\langle v_{i} \left(\mathbf{x}\right)v_{j} \left(\mathbf{x}\right)\omega _{k} \left(\mathbf{x}+\mathbf{r}\right)\right\rangle $ in the same frequency range as for the helicity components, the spectrum slope close to -1 is observed (Fig. \ref{Fig3}).

\begin{figure}
 \includegraphics[width=1.0\textwidth]{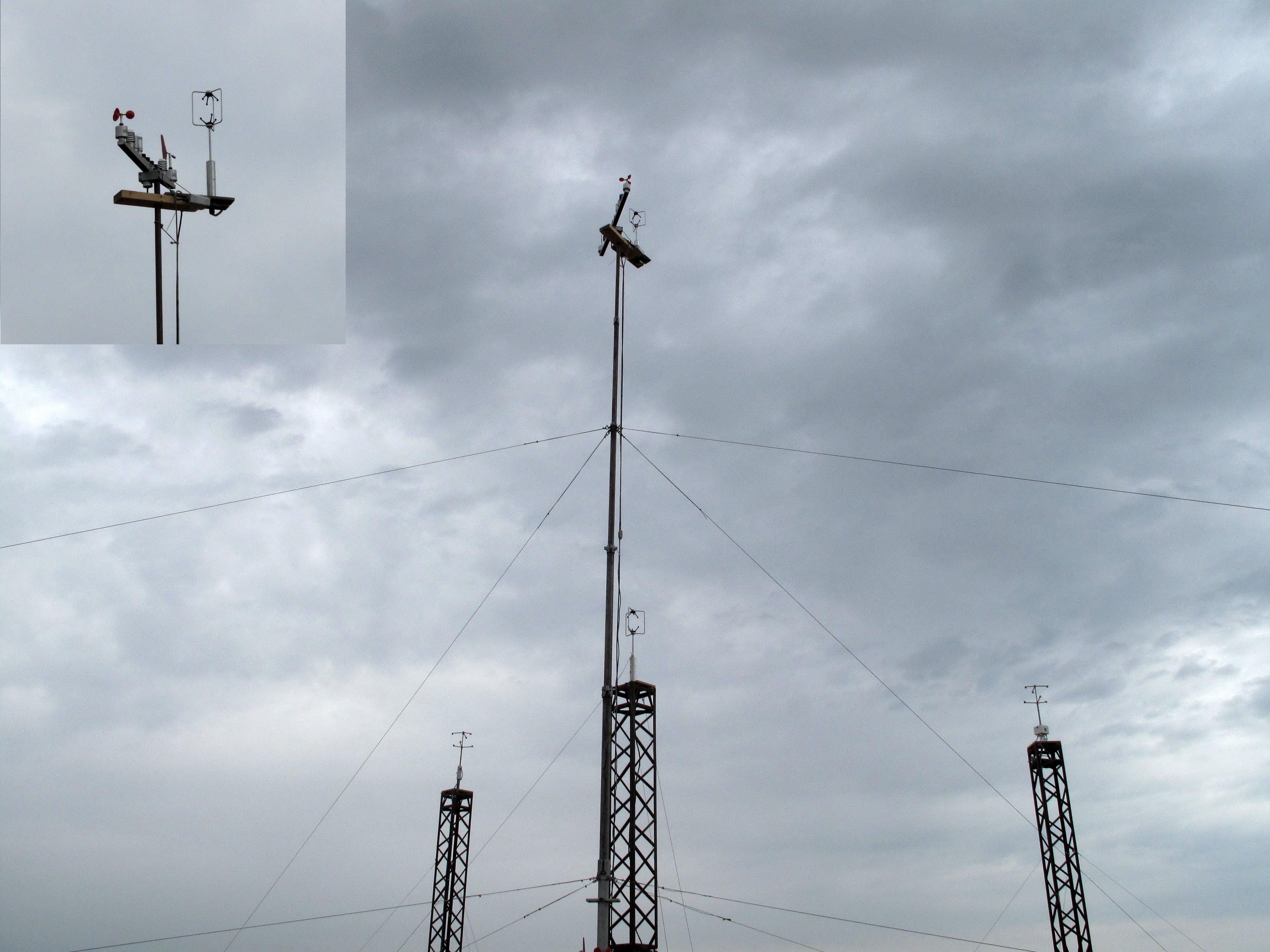}
\caption{Photo of the tetrahedron composed of four sonic anemometers to measure helicity in \citet{Koprov2015}; see also \citet{Chkhetiani2018}. An enlarged image of the uppermost anemometer at 10-m height assembled with a cup anemometer and a windvane (temperature, pressure and humidity sensors) is in the upper left corner. (Photo copyright O. Chkhetiani)
\label{Fig4}}
\end{figure}

\begin{figure}
 \includegraphics[width=1.0\textwidth]{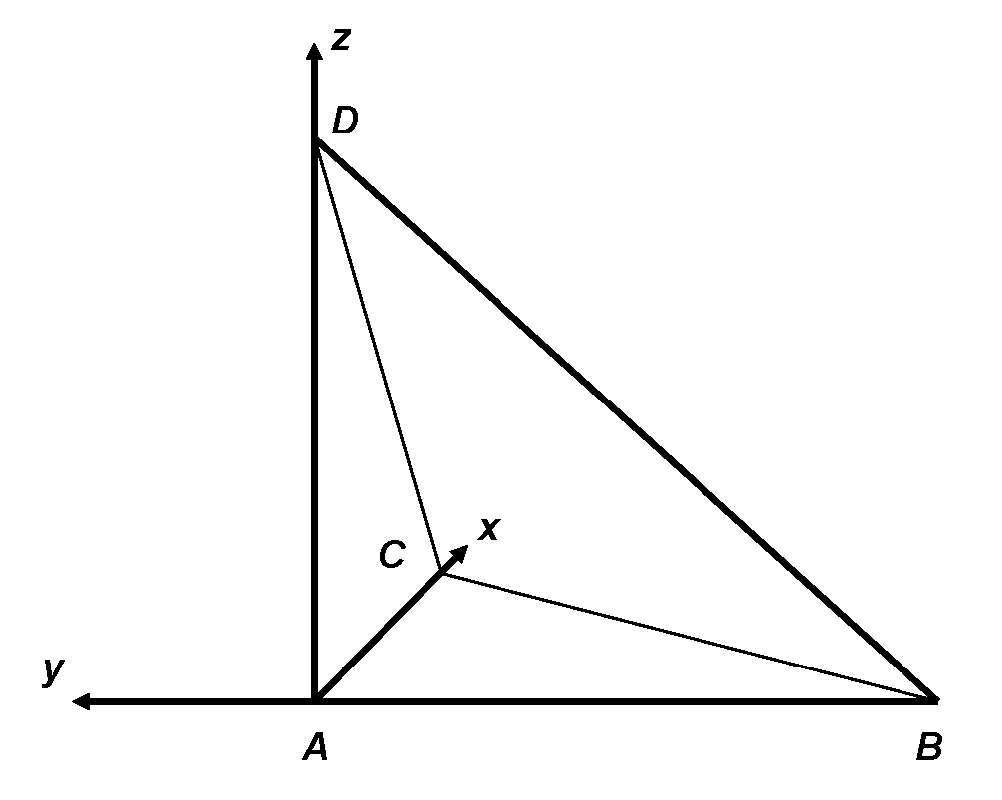}
\caption{Schematic of the rectangular tetrahedron \textit{ABCD} used for helicity measurements; the (\textit{x}, \textit{y}) axes are directed north and west, respectively. Three anemometers are positioned at the vertices of a rectangular isosceles triangle \textit{ABC} with sides \textit{AB} and \textit{AC} equal to 5 m. The fourth anemometer at point \textit{D} is located 5 m higher, just above poimnt \textit{A}; see also \citet{Chkhetiani2018}.
\label{Fig5}}
\end{figure}

In a field campaign in Tsimlyansk, Russia in August 2012,  \citet{Koprov2015} use four three-component acoustic anemometers in the vertices of a rectangular tetrahedron with the side of 5 m and its base at 5.5 m above the ground (Fig. \ref{Fig4}). Two different methods of calculating helicity were applied. First, the velocity circulations along the contours encircling the tetrahedron faces were calculated and then the Stokes theorem was applied to calculate the corresponding average vorticities, $\omega _{x} $, $\omega _{y} $ and $\omega _{z} $, and thus the helicity. Second, an alternative finite-difference method for approximating,  $\omega _{x} $, $\omega _{y} $ and $\omega _{z} $  in point \textit{A} (see Fig. \ref{Fig5})was applied to calculate helicity, see \eqref{Eqn_15},
\[
H_{A} =u_{A} \left(\omega _{x} \right)_{A} +v_{A} \left(\omega _{y} \right)_{A} +w_{A} \left(\omega _{z} \right)_{A}
\]
\begin{equation} \label{Eqn_15}
\approx \frac{w_{A} u_{B} -u_{A} w_{B} }{L_{AB} } +\frac{v_{A} u_{D} -u_{A} v_{D} }{L_{AD} } +\frac{w_{A} v_{C} -v_{A} w_{C} }{L_{AC} } .
\end{equation}
The three components $\left(u,v,w\right)$  of velocity are measured in points \emph{A}, \emph{B}, \emph{C} and \emph{D}; $L_{AB}$, $L_{AD}$ and $L_{AC}$ are the lengths of the tetrahedron edges, see Fig. \ref{Fig5}. Both methods give similar results.

Turbulent helicity measured values are shown in Table \ref{tabl-2}. Primes denote deviations from the values, obtained by running mean averaging with a 30-seconds rectangular window; $\theta '$ is the temperature fluctuation. Hereafter, an overbar denotes averaging over the entire measurement session. More details are given in \citet{Chkhetiani2018}.

\begin{table}
\caption{The values of co-variances obtained experimentally in \citet{Koprov2015}, see also \citet{Chkhetiani2018}. The  $x$-axis is directed north, the  $y$-axis west, and the  $z$-axis up. Values in the middle row correspond to the measurement session from 12:38 to 16:30 local time (UTC + 3 h) on 8 August 2012. The numbers in the lower row correspond to the measurement session from 22:46 to 00:34 local time, the end of the session on 9 August 2012.\label{tabl-2}}
{
\begin{tabular}{@{}cccccc@{}}
\toprule
& $\overline{u'\omega '_{x} }$, ms${}^{-}$${}^{2}$
& $\overline{v'\omega '_{y} }$, ms${}^{-}$${}^{2}$
& $\overline{w'\omega '_{z} }$, ms${}^{-}$${}^{2}$
& $\overline{w'\theta '}$, mKs${}^{-}$${}^{1}$
& $\overline{\omega '_{z} \theta '}$, Ks${}^{-}$${}^{1}$ \\
\midrule
Day & --0.0170 & --0.0276 & 0.0143 & 0.1545 & 0.0456 \\
Night & --0.0138 & --0.0010 & 0.0085 & --0.0172 & --0.0123 \\
\botrule
\end{tabular}
}
{}
\end{table}

\begin{figure}
 \includegraphics[width=1.0\textwidth]{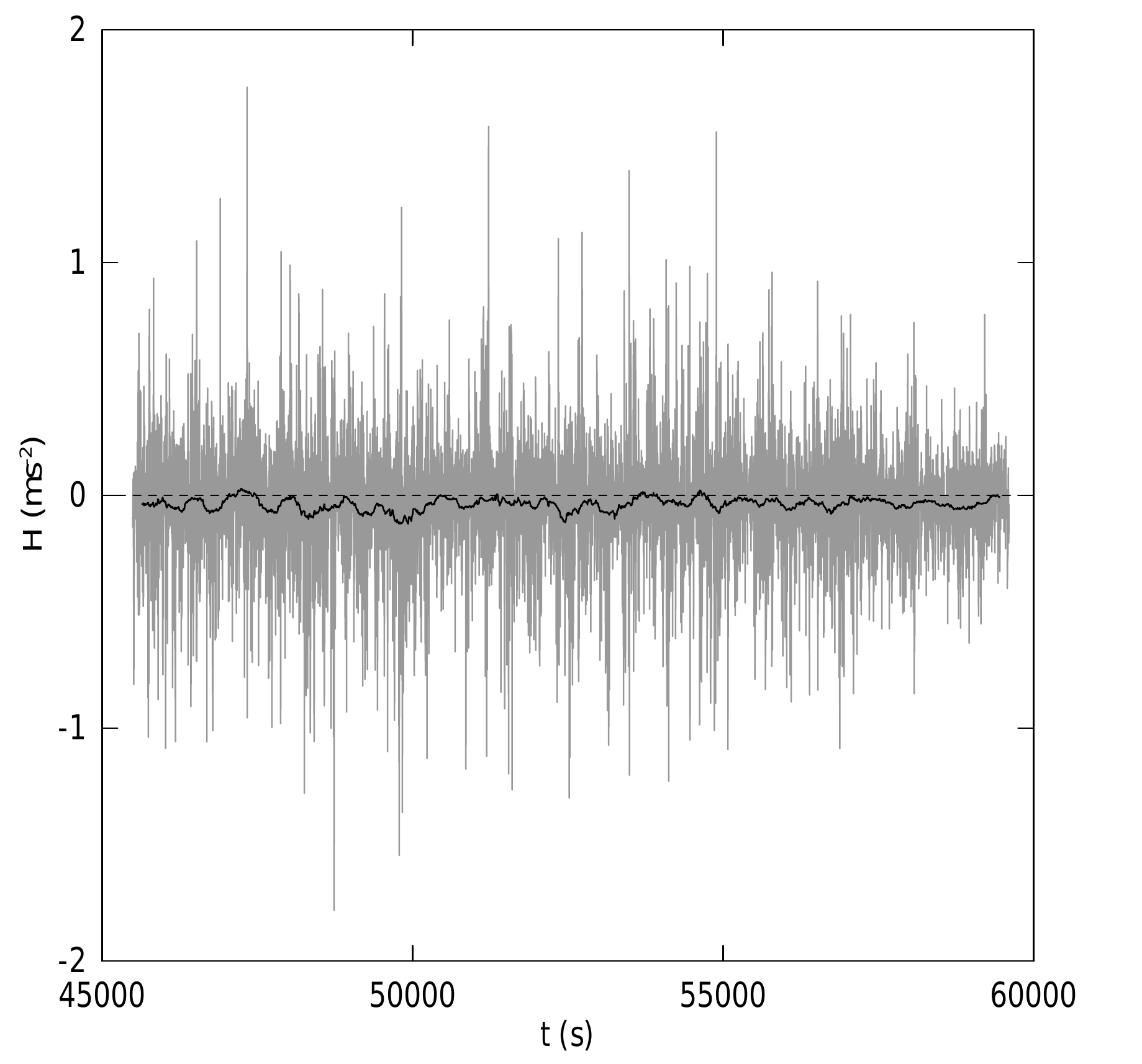}
\caption{Turbulent helicity $H'$ values (in ms$^{-2}$) measured during the day in \citet{Koprov2015} and averaged over 1 sec (thin grey line); see Table \ref{tabl-2} and \citet{Chkhetiani2018}. The abscissa shows the elapsed time (in seconds) since the beginning of the day. A solid line shows 5-min averages of  $H'$. \label{Fig6}}
\end{figure}

We emphasize that for the negative mean helicity the instantaneous helicity values can be of any sign and the negative ones dominate only in a statistical sense, and the corresponding standard deviation is significantly higher than the mean value, see Fig. \ref{Fig6}.

\begin{figure}
 \includegraphics[width=1.0\textwidth]{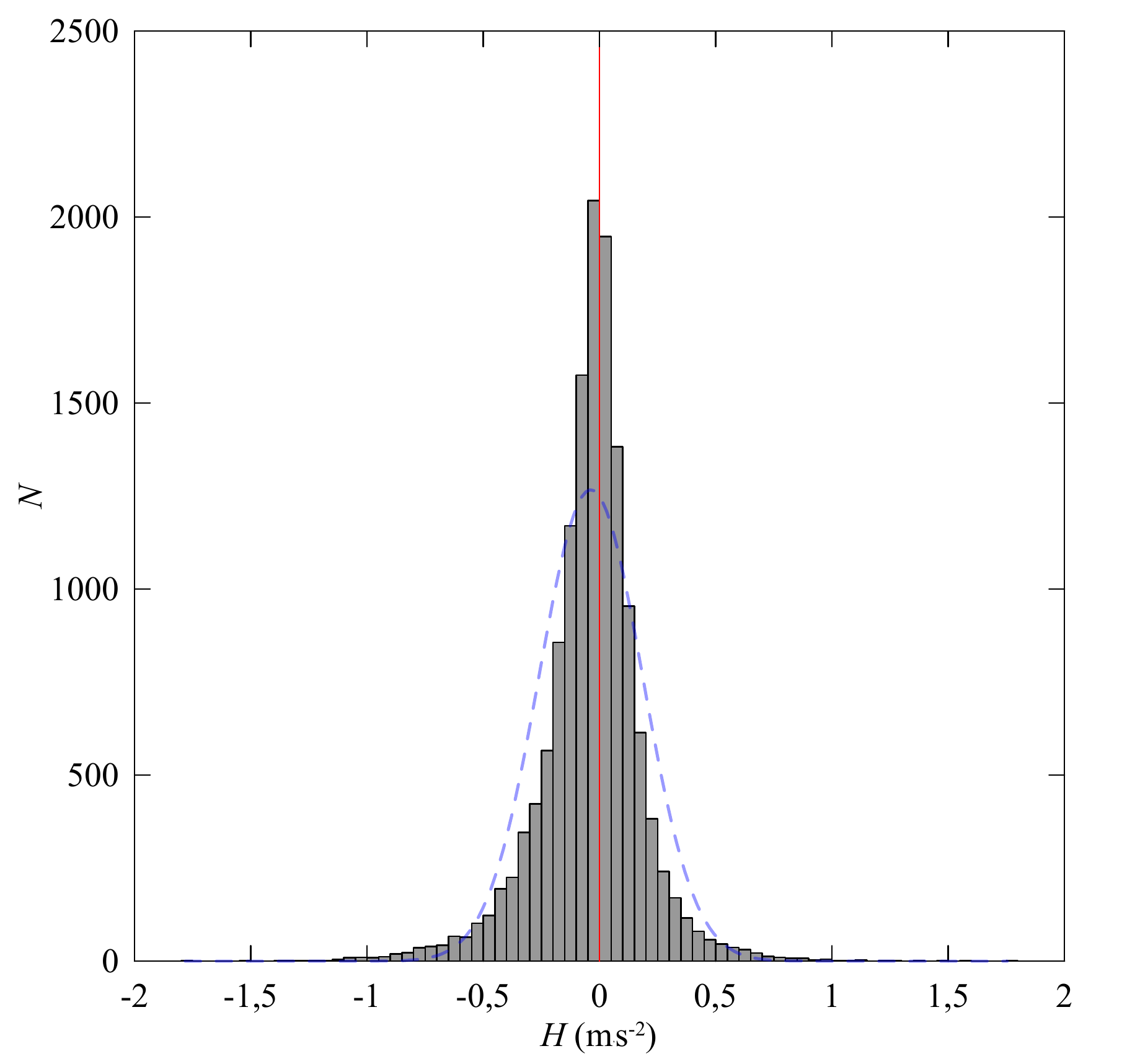}
\caption{Histogram of instantaneous turbulent helicity values measured in Tsimlyansk, southern Russia on 8 August  2012 at 12:38$-$16:30 local time (\citet{Koprov2015}; see the text for more details). On the horizontal axis the turbulent helicity values (in ms${}^{-}$${}^{2}$) are shown; the width of each bin equals 0.05 ms${}^{-}$${}^{2}$. On the vertical axis, the numbers show the number \textit{N} of helicity values falling within the corresponding bin. The sum of all \textit{N} values equals 14,113 (see the text). A vertical red line shows zero helicity values \textit{H }$=$ $0$. For comparison, a Gaussian distribution that has the same dispersion is shown by a blue dashed line.
\label{Fig7}}
\end{figure}

Figure Fig. \ref{Fig7} shows the histogram of turbulent helicity $H'$ values presented  Fig. \ref{Fig6}. The measurement session lasted 14,113 seconds, therefore dividing the $N$ values in Fig. \ref{Fig7} by 14,113 we infer the probability density function of $H'$ values, which can be compared with Fig. 7 in \citet{Deusebio2014} for an idealized Ekman layer in the Northern Hemisphere, where, in contrast, positive turbulent helicity values dominate.

We now proceed to compare the experimental results by \citet{Koprov2015} with theoretical estimates following the idealized conceptual scheme by \citet{Deusebio2014}. More details are given in \citet{Chkhetiani2018}. Here, only the main stages of reasoning are described. In the Northern Hemisphere, positive helicity is injected from the free atmosphere into the Ekman boundary layer with the time rate $f\mathbf{u}_{g}^{2}$, where  $\mathbf{u}_{g}$ is the vector of geostrophic wind. Furthermore, helicity is cascaded as an effective passive scalar quantity towards smaller scales to be finally destroyed by viscosity. Then it follows that the helicity destruction rate in the lower part of the Ekman boundary layer reads
\begin{equation}\label{Eqn_16add}
\eta=2f\mathbf{u}_{g}^{2}/h_E,
\end{equation}
where $h_E=\sqrt{2K/f}$  is the Ekman boundary layer thickness, $K$ is the small-scale eddy viscosity, $f$ the Coriolis parameter, and the factor ''two'' is derived from more detailed considerations of helicity and superhelicity of the classical Ekman spiral solution in \citet{Kurgansky1989}: see more in \citet{Chkhetiani2018}.
Turbulent helicity $H'\equiv \overline{{\mathbf u'}\cdot {\bm \omega '}}=H'\left(l\right)$ on the
given spatial scale $l$  that belongs to an inertial interval can be estimated from \citet{Brissaud1973},
\begin{equation} \label{Eqn_16}
H^{t} \left(l\right)=C_{H} \eta \, \varepsilon ^{{-1/3} } l^{{2/3} } ,
\end{equation}
which corresponds to the joint cascade of energy and helicity to small scales to cf. \eqref{Eqn_9}; see also \citet{Chen2003}. In \eqref{Eqn_16},   $C_H$ has been estimated either as 2.3 \citep{Andre1977}  or 2.5 \citep{Avinash2006}.  Taking $h_{E} =400\, {\rm m}$, $f=10^{-4} {\rm s}^{-1} $, $\left|u_{g} \right|=10\, \, {\rm m}\, {\rm s}^{-1} $, $\varepsilon =10^{-3} \, {\rm m}^{2} {\rm s}^{-3} $, $C_{H} =2.4$, and $l\, =10\, {\rm m}$, cf. \citet{Koprov2005}, we obtain $H^{t} \approx 5.5\times 10^{-3} \, {\rm m}\, {\rm s}^{-2} $, that is, several times smaller than the daytime $H^{t} $ values measured in \citet{Koprov2005}.

In fact, a lower estimate was for $H^{t}$, which ignores other possible sources of turbulent helicity in the bottom part of the planetary boundary layer. \citet{Chkhetiani2018} discuss these and other possible complicating factors. The analysis of meteorological conditions during the field campaign in \citet{Koprov2015} suggests that the joint action of the Ekman spiral flow and the sea-breeze circulation explains the resulting negative sign of $H'$ and its magnitude, see Table \ref{tabl-2} and \citet{Chkhetiani2018}. This hypothesis was tested using the balance model of $H'$  in the planetary boundary layer \citep{Chkhetiani2018}.

It is also mentioned in \citet{Chkhetiani2018} that the scale $\lambda _{D} \sim \varepsilon ^{{2/7} } \nu ^{{3/7} } \left({2fu_{g}^{2}/h_{E} } \right)^{-{3/7} } $ obtained by equating \eqref{Eqn_16add} and $\eta =2\nu\overline{{\bm \omega }'\cdot \nabla \times {\bm \omega }'}$ corresponds to the Ditlevsen scale (\citet{Ditlevsen2001a,Ditlevsen2001b}; see also \citet{Ditlevsen2010}). Here, $\nu $ is the air molecular viscosity. For the parameter values pertinent to the measurement site in Tsymlyansk $\lambda _{D} \sim 1.1\times 10^{-1} {\rm m}$ is two orders of magnitude greater than the Kolmogorov microscale of turbulence, which is the minimum scale of helical motions for a joint cascade of energy and helicity (\citet{Chen2003}; see also, \citet{Briard2017}).

For a thermally stratified boundary layer the turbulent helicity reads  \citep{Chkhetiani2018}
\begin{equation} \label{Eqn_18}
H^{t} =2K\tau \left(\frac{d^{2} U}{dz^{2} } \frac{dV}{dz} -\frac{d^{2} V}{dz^{2} } \frac{dU}{dz} \right)+f\frac{d\, \overline{w'^{2} }}{dz} \, \tau +2\alpha g\overline{\omega '_{z} \theta '}\, \tau .
\end{equation}
where $\alpha>0$  is the coefficient of thermal expansion of air and $\tau$  the turbulent relaxation time. There are three different terms in the expression for  $H'$. The first term corresponds mainly to $\overline{u'\omega '_{x} }+\overline{v'\omega '_{y}}$ and an expression in
parentheses represents the superhelicity  $S$ of the mean flow. In the absence of the other two terms in \eqref{Eqn_18}, this term denotes $H'<0$  for $S<0$. The second term, generally positive in the Northern Hemisphere is two or three orders of magnitude smaller than the first term. The third term is due to thermal stratification and corresponds to the three contributions to helicity, including $\overline{w'\omega '_{z} }$. As the data in Table \ref{tabl-2} show, it is positive during the day and negative at night, respectively.

For the third term in \eqref{Eqn_18} we have in a steady statistically equilibrium turbulent regime
\begin{equation} \label{Eqn_19}
\tau ^{-1} \overline{\omega '_{z} \theta '}\approx \overline{w'\left(\overline{{\bm\omega}}\cdot \nabla \theta '\right)}+\overline{w'\left({\bm\omega}'\cdot \nabla \theta '\right)}.
\end{equation}

\begin{figure}
 \includegraphics[width=1.0\textwidth]{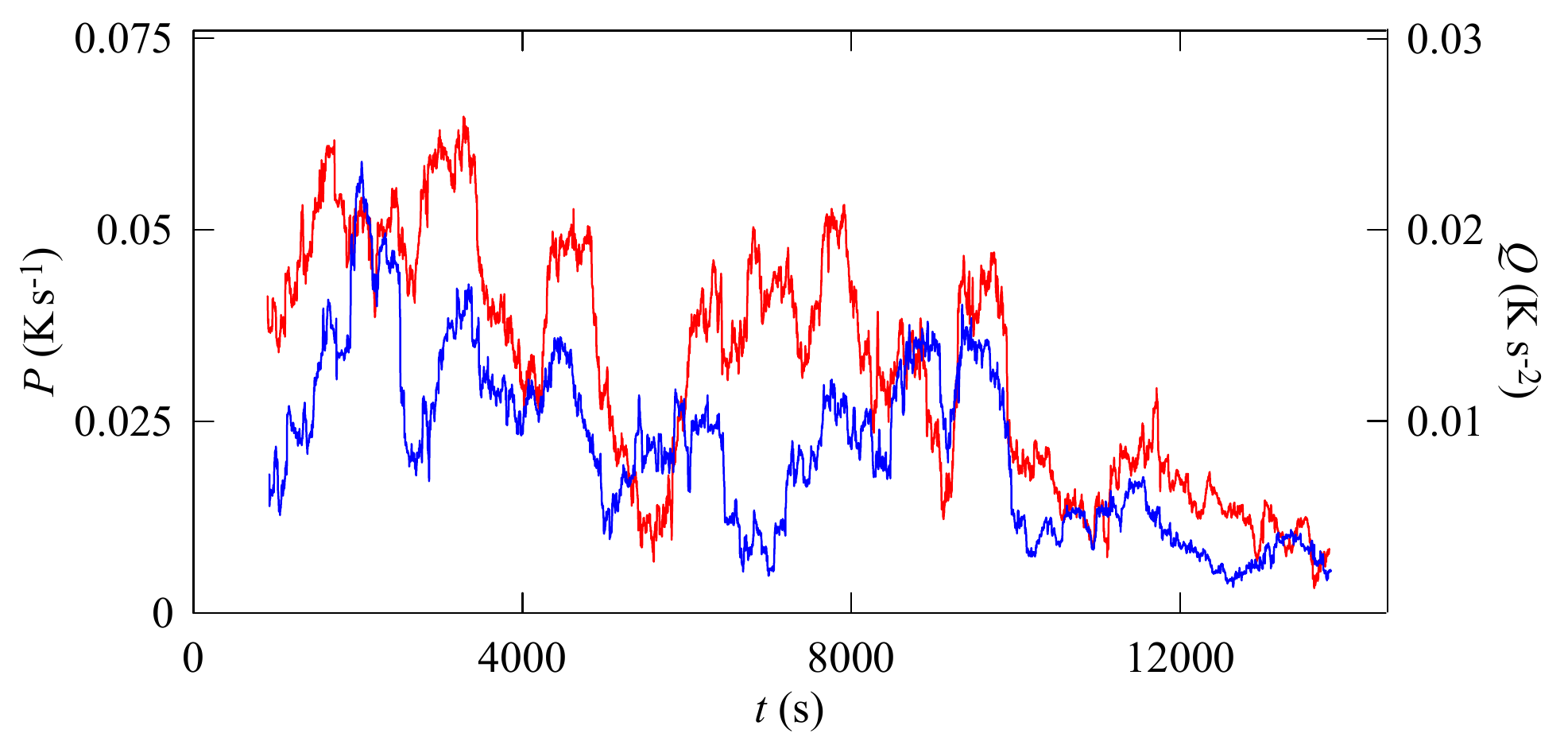}
\caption{Time dependence of $P$ (in Ks${}^{-}$${}^{1}$, red line) and $Q$ (in Ks${}^{-}$${}^{2}$, blue line) values, for the daytime measurements in \citet{Koprov2015}, see also Table \ref{tabl-2} and the text for details. On the horizontal axis, zero corresponds to 12:48 local time \citep{Chkhetiani2018}.
\label{Fig8}}
\end{figure}

It should be noted that the right-hand side of \eqref{Eqn_19} is nothing but a mutual correlation of the vorticity fluctuations and the turbulent Ertel potential vorticity. In Fig. \ref{Fig8} we show the time dependence of two fluctuating quantities, $P\equiv \omega '_{z} \theta '$  and   $Q\equiv \left(w'\left(\overline{{\bm\omega}}\cdot \nabla \theta '\right)+w'\left({\bm\omega}'\cdot \nabla \theta '\right)\right)$, for the daytime measurements in \citet{Koprov2015}, see Table \ref{tabl-2}. A 10-min averaging is used to plot $P$  and $Q$, and the correlation between these quantities is evident in Fig. \ref{Fig8}. Analogous results are obtained for the nighttime measurements in \citet{Koprov2015}.

In 2014, measurements were carried out using a similar scheme with a basic setup scale of 0.7 m at heights of 3.5, 13.1, and 25 meters \citep{Koprov2018}. Differences from zero of all three diagonal elements of the vorticity flux tensor, the correlation of the vorticity with the temperature field, and spectral dependences for helicity were found, close to those following from the concept of the constancy of the helicity flux towards small scales in the inertial interval.

\section{Concluding remarks and outlook}\label{sec5}

The role and significance of the concept of helicity in atmospheric physics will be fully appreciated in the future. Among other things, this Chapter discusses the relationship between the concepts of helicity and potential vorticity, which is especially non-trivial in a compressible fluid. There is a certain complementarity of the concepts of potential vorticity and helicity. The first of them, historically at least, was mainly demanded for the ``clarification'' of the principles of the dynamics of large-scale processes. At the same time, the concept of helicity for such motions is not very significant (meaningful). The situation changes as the transition to meso- and then small-scale atmospheric motions occurs. It is not to say that the concept of potential vorticity is completely losing its meaning. In a sense, it becomes less constructive, since the ``invertibility principle'' \citep{Hoskins1985}, that is, the possibility of restoring all other (balanced) hydrodynamic fields from the known field of a potential vorticity, gradually loses its applicability. For large-scale motions, this possibility is guaranteed by the quasi-geostrophic theory and its improvements. For meso- and small-scale motions, considerations of symmetry (axial, etc.) are indispensable. At the same time, the helicity concept acquires more weight and its alignment occurs in the conceptual sense with the potential vorticity, which makes its connection with the latter more important.

As for the study of helical turbulence, which is the central theme of the second part of this Chapter, it must be said with certainty that it is a very prospective field of research. Three aspects related to the problems pending their complete solution can be highlighted. First, it is a tricky question of the sign of helicity, since unlike energy or enstrophy, helicity is not a positive definite quantity. In particular, the results presented in this Chapter show that the sign of turbulent helicity inferred in direct field measurements within the atmospheric boundary layer is highly dependent on the local physical and meteorological conditions at the observational site and from its geographical position. Therefore, it is highly desirable to conduct further research in this direction. In particular, this refers to helicity field measurements in the Southern Hemisphere. Second, there is some hope that, with the invocation of the helicity concept, we will be able to get closer to understanding the enigmatic problem of intermittence of atmospheric (geophysical) turbulence, but much future work is needed in this direction. Third, and what is also important and somewhat related to the two previous aspects, is the issue of the inverse cascade of energy towards larger scales in three-dimensional helical turbulence, which could help to explain the genesis and maintenance of meso- and small-scale vortex structures and to further clarify the role and even the need for general rotation in this process.

 \textbf{Acknowledgements }The authors deeply honor the memory of Raymond Hide, Semen Moiseev and Boris Koprov, who laid many of the ideas underlying this review and made a great contribution to the study of helicity. We thank Peter Ditlevsen and an anonymous reviewer for their valuable comments and suggestions that improved the paper.


\bibliographystyle{apalike}

\end{document}